\documentclass[12pt,preprint]{aastex}

\def\ltsima{$\; \buildrel < \over \sim \;$}
\def\gtsima{$\; \buildrel > \over \sim \;$}
\def\lsim{\lower.5ex\hbox{\ltsima}}
\def\gsim{\lower.5ex\hbox{\gtsima}}
\def\lapp{\ifmmode\stackrel{<}{_{\sim}}\else$\stackrel{<}{_{\sim}}$\fi}
\def\gapp{\ifmmode\stackrel{>}{_{\sim}}\else$\stackrel{<}{_{\sim}}$\fi}

\newdimen\minuswidth    
\setbox0=\hbox{$-$}
\minuswidth=\wd0
\catcode`@=\active
\def@{\kern\minuswidth}
\setbox0=\hbox{\rm0}
 
\shorttitle{The age of Bulge clusters}
\shortauthors{Origlia et al.}
 
\begin{document} 
\title{
Probing the Galactic Bulge with deep Adaptive Optics imaging: the age of NGC 6440\altaffilmark{1} }

\author{
L. Origlia\altaffilmark{2},
S. Lena\altaffilmark{3},
E. Diolaiti\altaffilmark{2},
F.R. Ferraro\altaffilmark{3},
E. Valenti\altaffilmark{4},
S. Fabbri\altaffilmark{3},
G. Beccari\altaffilmark{3}
}

\altaffiltext{1}{Based on   observations collected
at the European Southern Observatory (programs 075.D-0359(A) and 077.D-0316(A)),
at the VLT, Paranal and program 081.D-0371(A) at NTT, La Silla, Chile.}
\altaffiltext{2}  {INAF - Osservatorio Astronomico    
di Bologna, via Ranzani 1, I--40127 Bologna, Italy, livia.origlia@oabo.inaf.it, emiliano.diolaiti@oabo.inaf.it, 
giacomo.beccari@oabo.inaf.it}
\altaffiltext{3} {Dipartimento di Astronomia, Universit\`a degli Studi
di Bologna, via Ranzani 1, I--40127 Bologna, Italy, sebastiano.lena@studio.unibo.it, francesco.ferraro3@unibo.it, 
sara.fabbri5@studio.unibo.it}
\altaffiltext{4} {European Southern Observatory,
 Alonso de Cordova 3107, Vitacura, Santiago, Chile, evalenti@eso.org}
 
%

\begin{abstract}
We present first results of a pilot project aimed at exploiting the potentiality of 
ground based adaptive optics imaging 
in the near infrared to determine the age of stellar clusters 
in the Galactic Bulge.  
We have used a combination of high resolution adaptive optics (ESO-VLT NAOS-CONICA)
and wide-field (ESO-NTT-SOFI) photometry of the 
metal rich  globular cluster NGC 6440 located towards the inner Bulge, 
to compute a deep color magnitude diagram from the tip of the Red Giant Branch  down to
$J\sim 22$,  two magnitudes below the Main Sequence Turn Off (TO).
The magnitude difference between the TO level and the red Horizontal Branch has been
used as an age indicator. 
It is the first time that such a measurement for a bulge globular cluster 
has been obtained with a ground based telescope.
From a
direct comparison with  47 Tuc  and with a set of theoretical isochrones, 
we concluded that NGC 6440 is old and likely coeval to 47 Tuc.
This result adds a new evidence that the Galactic 
Bulge is $\approx$2 Gyr younger at most than the pristine, 
metal poor population of the Galactic Halo.
 
\end{abstract}
 

\section{Introduction}
Galactic globular clusters (GCs) are known since a long time to be the oldest
stellar population of our Galaxy, hence their accurate age determination
is crucial to settle the epoch of the Galaxy formation and a
lower limit to the age of the Universe.
 A major effort has been devoted to obtain
reliable ages for the Halo GC system \citep{ros99,ros00,sw02,dea05}.

With the growing awareness that the Bulge GCs are a
distinct sub-system, has come a new urgency to properly characterize their
evolutionary sequences and to determine their ages and chemical composition. 
Moreover, Bulge GCs are {\it  key templates} of simple stellar populations  
to study the
stellar and chemical  evolution in the high metallicity domain \citep{mcw97,mat99,wys00,bal07} 
and  for population synthesis of giant elliptical galaxies.   
Hence, a proper characterization of
the main chemical and evolutionary properties of the Galactic Bulge and its 
comparison with the other components of the Galaxy is a
fundamental step to unveiling the process of Galaxy formation and evolution \citep{dut02,wys00}.

Despite its importance,  the Galactic Bulge GC system 
remained mostly unexplored, because of the huge foreground extinction, 
which severely prevents the usage of optical observations.
Also, the limited  photometric
and spectroscopic performances of the previous generations of infrared (IR) arrays, 
prevented  any deep, high resolution imaging and spectroscopy to 
properly survey the Bulge field and GC population. 

Over the last decade a few optical and near IR ground-based
photometric studies of bulge  GCs have been performed \citep{ort96,gua98,fro95,dav00}. 
Recently, \citet{val07} presented homogeneous and accurate IR color-magnitude diagrams 
(CMDs) of the evolved 
sequences of a sample of 24 GCs in the Bulge direction.  

However, only a few attempts exist so far to estimate the age of  Bulge
GCs from the direct measurement of the Main Sequence (MS) TO,  
by using the
WFPC2 and NICMOS onboard HST 
\citep{ort95,hea00,ort01,zoc01,coh02,fel02}. 

The recent advent of adaptive optics (AO) capabilities on 8m-class telescopes allows unprecedent
deep ground-based IR photometry down to the TO level in clusters as distant as
$\approx$10 kpc with potential better spatial sampling and coverage than NICMOS
onboard HST.    We performed a pilot project aimed at measuring   the age of the
Bulge GCs, by taking advantage of the ESO-VLT NAOS-CONICA (NACO)  imaging facility
with AO and wavefront sensing in the IR, which is crucial  to find natural guide stars
bright enough in such a  reddened environment. In this Letter, we present the first
results of such a project, namely the age determination of the  GC NGC 6440.  This
cluster has an iron abundance [Fe/H]=-0.56 \citep{ori08} and a global metallicity
[M/H]=-0.4  \citep{val07} and it is located in the inner Bulge at (l,b)=(7.7,3.8)
and 8.2 kpc from the Sun.  The high reddening \citep[E(B-V)=1.15, ][]{val07}
along the line of sight yields an extinction of $\approx$3.5 mag 
in the visible, and makes this cluster an ideal target to
be investigated in the near IR.

\section{Observations and Data Reduction}

\begin{figure}[!hp]
\begin{center}
\includegraphics[scale=0.7]{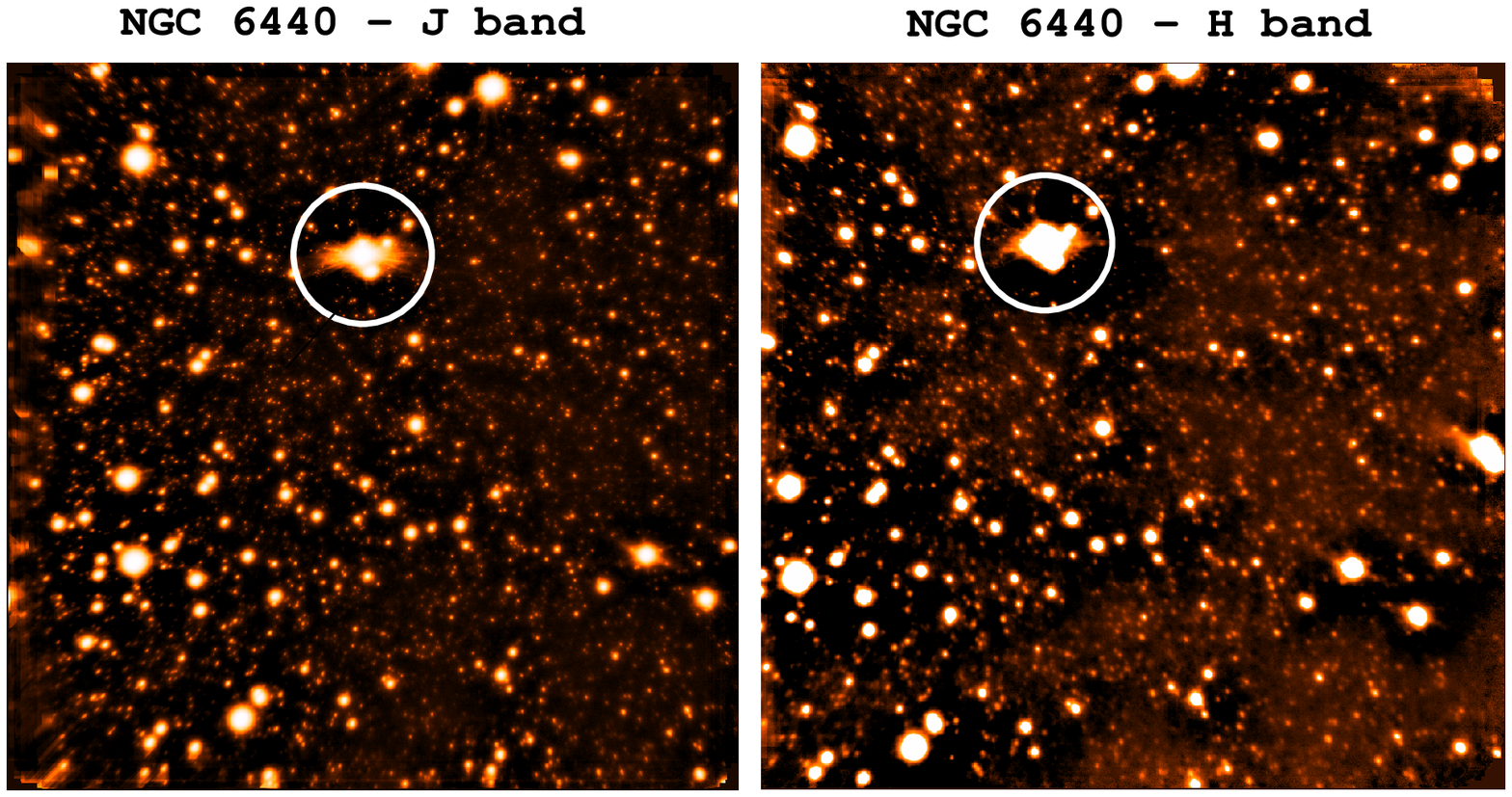}
\caption{NACO J and H band images of the observed field ($\approx$50''$\times$50'') 
located $\approx$110'' South-West from the NGC 6440 center. 
North is up, East is left. The circled star is the natural guide star used for AO correction.}
\label{ima}
\end{center}
\end{figure}

The observations have been performed in service mode between  August 2005 and
September 2006 as part of the programmes 075.D-0359(A) and
077.D-0316(A) (PI: Origlia), by using
NACO mounted at the ESO-VLT.
We used the S54 camera which offers a relatively large
$55\arcsec\times 55\arcsec$ FoV with sufficient 
($0\farcs054\,{\rm pix}^{-1}$) spatial resolution to resolve the stars. 
We selected an appropriate 
field located at $\approx$110'' South-West from the cluster center,  where
crowding is not too severe but the 
star density is still large enough to sample
a reasonable number of  stars  along the evolved sequences of the CMD 
(like the Sub Giant Branch). 
A series of 24 exposures each one 90 
sec-long (DIT=18 sec and NDIT=5 and DIT=10 sec and NDIT=9, in the J and H
bands, respectively) have been secured. 
Fig.~\ref{ima} shows the NACO J and H mosaiced images, with marked 
the natural guide star used to perform the AO correction.

We used the final mosaiced images obtained from the reduction pipeline  and
computed PSF-fitting photometry by using STARFINDER \citep{dio00} specifically
developed and optimized to obtain accurate photometry in  AO-corrected crowded
fields. Indeed, STARFINDER allows the modeling of the PSF as a function of the 
distance from the  reference star with a combination of Gaussian/Moffat functions
to simultaneously account for the stellar peak and the  diffuse seeing halo.  
More specifically,  
the complex NACO-PSF in the NGC 6440 images has been modeled with a combination of
three Gaussians. Two Gaussians have been used to properly model the
central peak of the PSF (their parameters strongly depending  on the efficiency of
the AO correction), while the third Gaussian parametrizes the PSF halo, and  
depends on the seeing.  Since the AO corrections
degrade with the distance from the reference star, STARFINDER has been optimized 
to allow the parameters of the two Gaussians modeling the central peak to vary as a function of the
distance from the reference star. All the parameters of the Gaussians have been
fine-tuned by properly modeling a number of high signal-to-noise ratio, unsaturated stars
over the entire science image.  Once the PSF has been determined, an
automatic search of the stars has been performed and the PSF fitting
procedure executed for all the detected objects. At the end of this
procedure, a final list of instrumental magnitudes and coordinates for more than
2000 stars has been obtained. 

Complementary to the NACO observations of the TO region, 
we have also acquired shallower ($\approx$1min exposures) J,H,Ks images over a wider (5'$\times$5') field of view 
at the ESO-NTT using SOFI (programme 081.D-0371(A), PI Valenti) to 
sample the brightest portion of the CMD, namely the Red Giant Branch (RGB) and the Helium clump.
The SOFI J and H images have been reduced by using ROMAFOT \citep{buo83} 
and the instrumental magnitudes have been calibrated and placed  on the absolute astrometric system 
of 2MASS. The brightest, unsaturated stars in the NACO field which are in common 
with the SOFI dataset have thus been used to
transform the NACO J and H instrumental magnitudes to the   
SOFI/2MASS photometric system. 
The average photometric uncertainty turns out to be a few hundreths mag along the RGB and 
$\approx$0.1 mag in the TO region. 
From a direct comparison of our SOFI photometry of the cluster with  
a control field from 2MASS we find that field contamination 
along the  RGB is below 20\% in the NACO field of view. 
By performing a number of simulations with the \citet{rob03} 
model of stellar population synthesis of the Galaxy, 
we also estimate a similar degree of field contamination in the TO region.
Hence we conclude that such a degree of field contamination should not appreciably affect 
our estimates of the Horizontal Branch (HB) and TO levels.

As a template of an old, metal rich GC we took 47 Tuc ([Fe/H]=-0.67)
\citep{car00},  for which an accurate absolute age estimate ($11.2\pm1.1$ Gyr) has
been recently obtained by \citet{gra03}. As was done in the case of NGC 6440, we use
SOFI shallow photometry of the cluster center to sample the RGB \citep{ori07}  and
deep photometry of the TO region as obtained from  archival SOFI J,H images of a
field located at $\approx$2' from the cluster center \citep[see also][]{sal07}. 
The photometric analysis of this dataset was performed by using ROMAFOT
\citep{buo83}. The sample has been astrometrized and photometrically calibrated by
using the stars in common with the 2MASS catalog.

\section{Results}

\begin{figure}[!hp]
\begin{center}
\includegraphics[scale=0.7]{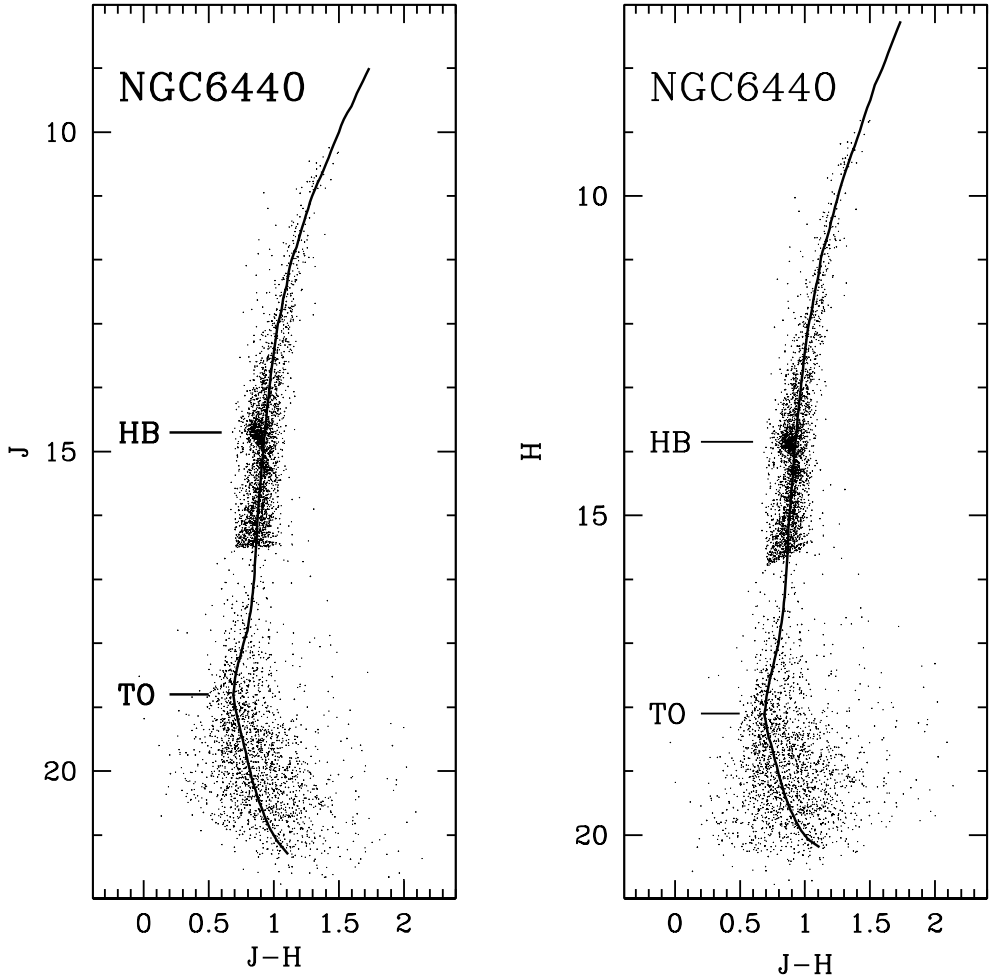}
\caption{Combined NACO-SOFI (J,J--H) and (H,J--H) CMDs
for NGC 6440. Only stars with $J<16.5$ from the SOFI sample and
with $r<30"$ from the reference star in the NACO sample are plotted.
The RGB-SGB-MS mean 
ridge lines  are overplotted to the data. 
The HB and TO levels are also marked for reference.}
\label{cmd}
\end{center}
\end{figure}

Fig.~\ref{cmd}   shows the NACO-SOFI $J,J-H$ and $H,J-H$ combined CMDs for NGC 6440. 
Only stars with $J<16.5$ in
the SOFI sample and within $30"$ from the reference star in the NACO sample
are plotted. All the
evolutionary sequences from the RGB tip down to the MS are sampled. In particular,
the CMDs reach two magnitudes below the
cluster TO and offer one of the cleanest insights  of the MS TO region ever
published for a cluster in the inner Galactic Bulge.    
These observations show that an AO capability at an 8m class telescope  is
crucial to obtain samples of stars
{\it i}) $\approx$1 mag fainter and {\it ii}) poorly affected by incompleteness due to crowding 
with respect to those obtained from average seeing limited observations. 
These samples allow us to properly 
measure the TO region even in severely obscured star clusters in the Galactic Bulge.

The mean ridge line of the RGB/SGB/MS
sequence has been determined by 
applying an iterative 3$\sigma$-clipping procedure to the observed points
and it is overplotted on the CMDs.

The obtained CMDs can be used to measure the age of NGC 6440. 
As has been well known for many years  \citep{ibe74,san86} the difference in magnitude
between the HB and the TO is a sensitive function of the cluster age for any
stellar population older than 2 Gyr. In fact in this age range  the luminosity of
the HB  remains nearly constant while the MS-TO becomes fainter and
fainter with increasing age.  The $\Delta_{HB-MSTO}$ parameter, defined in different
photometric bands, has been successfully used 
in the past to derive relative ages
of Halo \citep{buo89,ros99,cha96} and Bulge \citep{ort01} GCs.

In the following we compute this parameter for NGC 6440 and the reference cluster 
47 Tuc, in order 
to derive the relative ages of these two stellar systems.
 
\begin{center}
\begin{figure}[!p]
\includegraphics[scale=0.7]{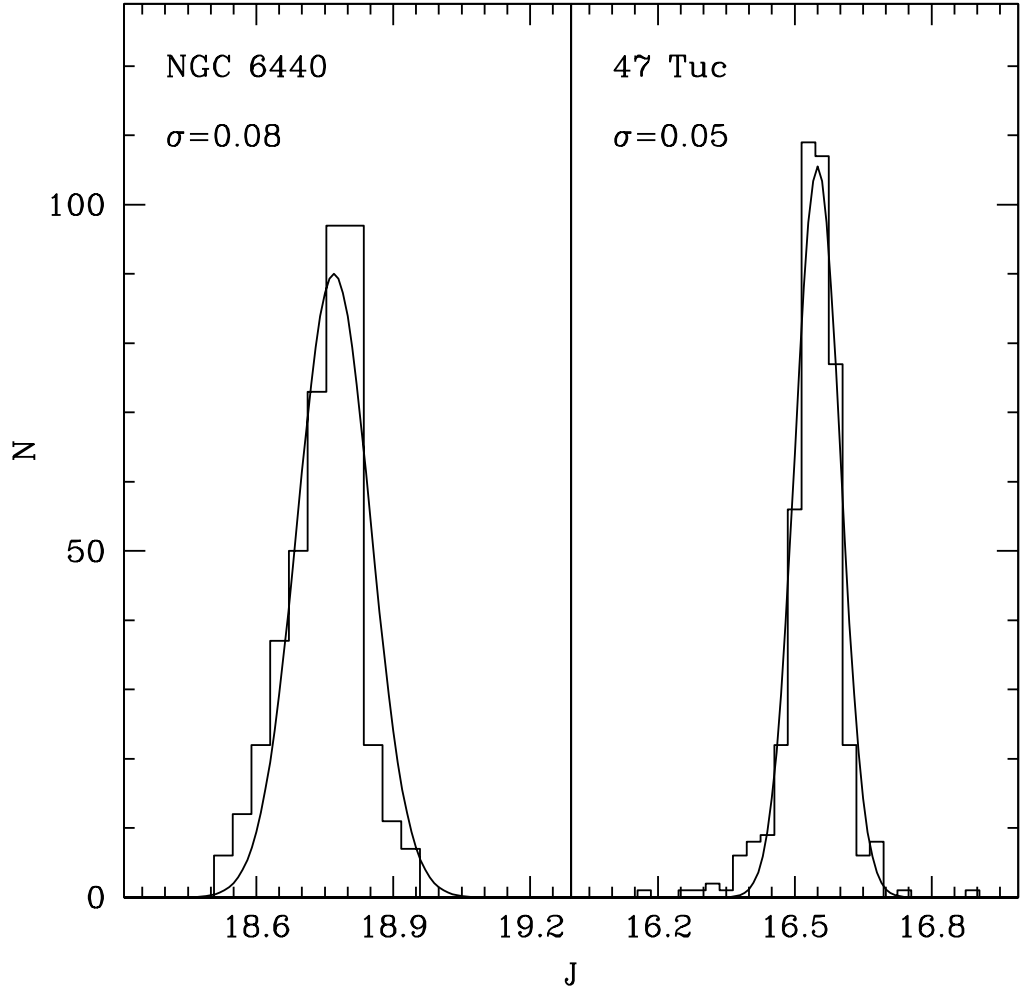}
\caption{Distribution of the TO level determinations as obtained from $\approx$ 400 fitting models 
of the TO region for NGC 6440 (left panel) and 47 Tuc (right panel). 
The Gaussian best fit and the corresponding $\sigma$ value for each distribution 
is also reported.
}
\label{dis}
\end{figure}
\end{center}

\begin{figure}[!hp]
\begin{center}
\includegraphics[scale=0.7]{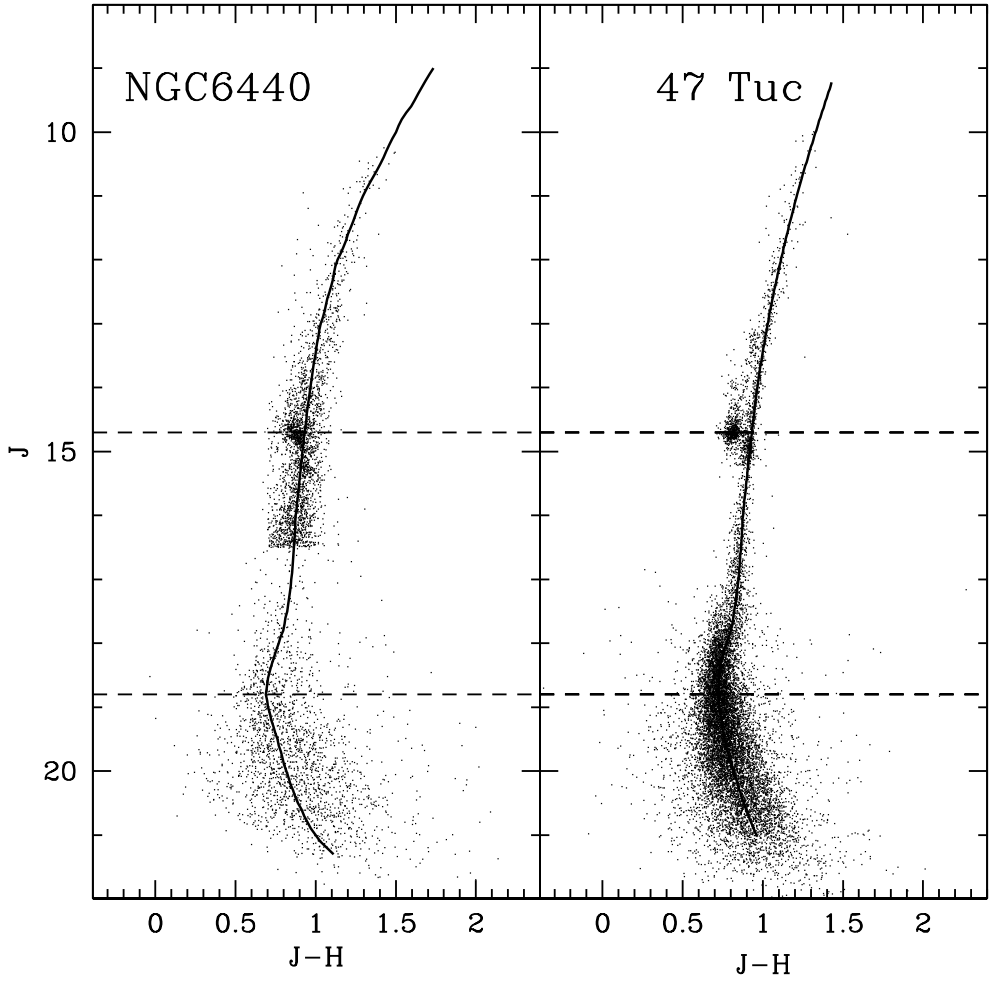}
\caption{The  $(J,J-H)$ CMDs
for NGC 6440 (left panel) and
47 Tuc (right panel), as shifted in color and magnitude, accordingly to the reddening and 
distance modulus of NGC 6440. 
The horizontal dashed lines mark the
HB and TO level of NGC 6440.}
\label{cmdconf}
\end{center}
\end{figure}

The HB levels in NGC 6440 turn out to be 
J(HB)=14.70 and H(HB)=13.85, respectively;  for comparison, 
47 Tuc shows J(HB)=12.5 and H(HB)=12.05. 
They have been computed by determining the peak of the HB Luminosity Function 
for stars  with $J<15$ and $(J-H)<0.95$.  
Such a selection allows a safe determination within $\pm 0.05$ mag  
of the mean level of the HB red  clump by reducing any possible 
contamination of RGB stars.
The MS TO points, defined as the bluest point along the mean ridge line 
of the RGB,SGB,MS are located at J(TO)=18.8 and H(TO)=18.1; 
for comparison, 47 Tuc shows J(TO)=16.55 and H(TO)=16.20. 
In order to estimate the errors in the TO level determination 
we follow \citet{mw06}.
We fitted the ridge line with a fifth-order (J-H)=f(J) polynomial 
and compute the $\rm \sigma(J_{TO})=\sigma(f(J))/\sqrt(N)$, 
where $\rm \sigma(f(J))$ is the standard deviation of f(J) 
within a $\rm 3\sigma(J-H)_{TO}$ color interval around the TO. 
The inferred $\rm \sigma(J_{TO})$ are 0.07 for NGC 6440 and 0.06 for 47 Tuc.
As an independent estimate we also compute the distribution of 
the $\rm J_{TO}$ values obtained by directly fitting the 
observed CMD TO region with a fifth-order polynomial, 
by varying the magnitude interval around the TO and the $\sigma$ rejection 
threshold between 2 and 4. 
The obtained distributions of the TO level determinations are plotted in Fig.~\ref{dis}. 
The best fit Gaussian $\sigma$ turns out to be 0.08 in NGC 6440 and 0.05 in 47 Tuc. 
We thus conservatively assume a $\pm 0.1$ mag error in NGC 6440 and $\pm 0.05$ mag in 47 Tuc.
On the basis of these measurements, the magnitude difference between 
the HB and the TO in the J band of NGC 6440 is $\rm \Delta J(HB-TO)=4.10\pm0.11$ 
and the corresponding value in 47 Tuc is $\rm \Delta J(HB-TO)=4.05\pm0.07$. 
These values suggest a similar age for the two clusters. 
The nice similarity between the CMDs of the two clusters is clearly 
visible in Fig.~\ref{cmdconf}, where the CMD of 47 Tuc (right panel) 
has been shifted in color and magnitude according to the reddening 
and distance of NGC 6440, and to match the HB level of the latter (left panel).
Note that the 
metallicity of NGC 6440 is slightly higher ($\approx$0.1 dex) than that of 47 Tuc.
However, inspection of the theoretical isochrones suggests that
a $\rm \delta[Fe/H]\approx0.1$ translates into a $\rm \delta J(HB-TO)\approx 0.03$, only, 
which is well within the errors. 
Hence, from this analysis we conclude that NGC 6440 is consistent with being coeval to 47 Tuc,
that is $\approx$11 Gyr old.

In order to evaluate the overall uncertainty of the $\Delta J(HB-TO)$ method 
in deriving absolute ages, we use theoretical isochrones of different ages. 
Two sets of $\alpha$-enhanced models at [M/H]=-0.65  and [M/H]=-0.35, 
respectively, to encompass the metallicity of 47 Tuc and NGC 6440, 
and ages ranging from 2 to 13 Gyr, have been retrieved from 
the BASTI database \citep{pie04} and transformed into the observational 
2MASS (J,J-H) plane following the prescriptions by \citet{ori00}. 
The difference between the mean value of the HB  and the TO level  
was computed for each isochrone\footnote{ 
As an independent test of the isochrones, we can also
use our measurement, we can also use our measurement of the  
$\rm \Delta J(HB-TO)$ in 47 Tuc and its robust age estimate by \citet{gra03} 
to evaluate the reliability of the theoretical $\rm \Delta J(HB-TO)$
in deriving absolute ages of a stellar system.
By using the [M/H]=-0.65 isochrone, we find that 47 Tuc turns out to be 
$\approx$10 Gyr old,
which is  $\approx$1 Gyr younger than in \citet{gra03}.
In this context, the $\rm \Delta J(HB-TO)$ value measured in NGC 6440 would
give ages $\approx$11 and $\approx$10 Gyr, by using 
the [M/H]=-0.65 and the [M/H]=-0.35 isochrones, respectively.}.
We find that in the metallicity range under consideration, 
a $\approx$0.1mag error in the $\rm \Delta J(HB-TO)$ value translates into 
a 2 Gyr uncertainty in the 8-11 Gyr age range and $\ge$3 Gyr between 10 and 13 Gyr, 
since $\rm \Delta J(HB-TO)$ becomes progressively less and less sensitive 
to age above 11 Gyr. 
Hence we can conclude that NGC-6440 is 11$^{+3}_{-2}$ Gyr, 
similar to 47 Tuc and other metal rich  GCs of the Galaxy.

\section{Discussion and Conclusions}

In this letter we have presented the age determination for NGC 6440: it is the first time that
ground based AO imaging in the near infrared has been used to successfully  measure the TO
region and derive the age of a Bulge GC. The few previous studies  have been all based on
space observations. In fact, since the pioneering work by \citet{ort95} who first measured
the TO region of NGC~6553 and NGC~6528  by using HST-WFC2 and suggested a near-coeval age
for the Galactic bulge and Halo,  a number of  subsequent works using HST and the WFPC2 and
NICMOS imagers  \citep{ort01,fel02} refined techniques and procedures to measure the age of
these two globular clusters,  by also applying proper motion decontamination to the CMDs
\citep{zoc01,fel02}. Only one other cluster in the inner Bulge, namely Terzan 5
\citep{ort01,coh02}  has been studied with HST-NICMOS, while \citet{hea00} analyzed
two other GCs observed with HST-WFPC2 in the outer Bulge, namely  NGC~6624 an NGC~6637.  
All these studies suggest that these clusters are old ($>$10 Gyr) and likely 
coeval with 47 Tuc.

While our and previous studies agree on the fact that Bulge GCs are likely coeval to 47 Tuc, 
absolute age estimates range between 11 and 14 Gyr, depending on the adopted
theoretical isochrones and/or the absolute age for 47 Tuc itself. In this respect, a
significant step forward in setting the absolute age of reference clusters   
has been performed by \citet{gra03} who provided accurate distances and absolute ages  for
three pillar GCs, namely NGC~6397, NGC~6752 and 47 Tuc, by using the MS Fitting Method and
the Hipparcos distance scale.  Their age estimate for 47 Tuc turns out to be 11.2$\pm$1 Gyr,
$\approx 2.6$ Gyr  younger than the other more metal poor Halo clusters.   A somewhat
younger age (by 1-2 Gyr) was previously suggested by \citet{ros99}, who also found  
slightly younger ages for other two metal rich clusters in their sample, namely NGC~6352 and
NGC~6838, which  possibly belong to the thick disk population, like 47 Tuc \citep[see also
Figure 2 in][]{fer04}.

These first set of age determinations seem to suggest a common  epoch of formation of the
tick disk and Bulge   and an overall GC formation process in our Galaxy starting $\approx
13.5$ Gyr ago \citep{gra03} with the formation of the  oldest Halo GCs and ending $\approx 11$ Gyr ago
with the formation of the metal rich objects in the thick disk and Bulge. In this scenario,
the few younger GCs found in the Halo (like for example Pal~12, Arp~2, Terzan~8 etc.) could not have
formed in situ. In fact, a number of recursive accretions of satellites and their GC
systems  \citep[as e.g. the Sagittarius Dwarf Spheroidal, ][]{iba94} could have
significantly contributed to form the present Halo stellar populations \citep{bel03,fer04}.

However, a larger sample of both Bulge and thick disk GCs urgently need 
to be observed with the purpose of determining their relative and absolute ages, before drawing firm conclusions.
This is something that is well within the capabilities of the current generation of instruments,  
such as the refurbished HST and the improving performances of ground based AO imagers.

\acknowledgements This research was supported by the {\it Ministero dell'Istruzione,
dell'Universit\'a e della Ricerca} and by the {\it Progetti Strategici 2006}
granted by the University of Bologna.
We thank the anonymous Referee for his/her helpful comments.

\clearpage

\end{document}